\documentclass[useAMS,usenatbib]{elsarticle}
\usepackage{hyperref}
\usepackage{graphicx}
\usepackage{amsmath,amssymb}

\newcommand{\f}{\frac}

\newcommand{\BF}{\begin{figure}\begin{center}}
\newcommand{\EF}{\end{center}\end{figure}}
\newcommand{\BE}{\begin{equation}}
\newcommand{\EE}{\end{equation}}
\newcommand{\BEA}{\begin{eqnarray}}
\newcommand{\EEA}{\end{eqnarray}}

\newcommand{\tr}{\textrm}

\newcommand{\ms}{M_{\odot}}
\newcommand{\vel}{\textbf{\textit{v}}}

\begin{document}
\def\aj{AJ}%
\def\actaa{Acta Astron.}%
\def\araa{ARA\&A}%
\def\apj{ApJ}%
\def\apjl{ApJ}%
\def\apjs{ApJS}%
\def\ao{Appl.~Opt.}%
\def\apss{Ap\&SS}%
\def\aap{A\&A}%
\def\aapr{A\&A~Rev.}%
\def\aaps{A\&AS}%
\def\azh{AZh}%
\def\baas{BAAS}%
\def\bac{Bull. astr. Inst. Czechosl.}%
\def\caa{Chinese Astron. Astrophys.}%
\def\cjaa{Chinese J. Astron. Astrophys.}%
\def\icarus{Icarus}%
\def\jcap{J. Cosmology Astropart. Phys.}%
\def\jrasc{JRASC}%
\def\mnras{MNRAS}%
\def\memras{MmRAS}%
\def\na{New A}%
\def\nar{New A Rev.}%
\def\pasa{PASA}%
\def\pra{Phys.~Rev.~A}%
\def\prb{Phys.~Rev.~B}%
\def\prc{Phys.~Rev.~C}%
\def\prd{Phys.~Rev.~D}%
\def\pre{Phys.~Rev.~E}%
\def\prl{Phys.~Rev.~Lett.}%
\def\pasp{PASP}%
\def\pasj{PASJ}%
\def\qjras{QJRAS}%
\def\rmxaa{Rev. Mexicana Astron. Astrofis.}%
\def\skytel{S\&T}%
\def\solphys{Sol.~Phys.}%
\def\sovast{Soviet~Ast.}%
\def\ssr{Space~Sci.~Rev.}%
\def\zap{ZAp}%
\def\nat{Nature}%
\def\iaucirc{IAU~Circ.}%
\def\aplett{Astrophys.~Lett.}%
\def\apspr{Astrophys.~Space~Phys.~Res.}%
\def\bain{Bull.~Astron.~Inst.~Netherlands}%
\def\fcp{Fund.~Cosmic~Phys.}%
\def\gca{Geochim.~Cosmochim.~Acta}%
\def\grl{Geophys.~Res.~Lett.}%
\def\jcp{J.~Chem.~Phys.}%
\def\jgr{J.~Geophys.~Res.}%
\def\jqsrt{J.~Quant.~Spec.~Radiat.~Transf.}%
\def\memsai{Mem.~Soc.~Astron.~Italiana}%
\def\nphysa{Nucl.~Phys.~A}%
\def\physrep{Phys.~Rep.}%
\def\physscr{Phys.~Scr}%
\def\planss{Planet.~Space~Sci.}%
\def\procspie{Proc.~SPIE}%

\begin{frontmatter}
\title{Detecting Sub-lunar Mass Compact Objects 
toward the Local Group Galaxies }
\author{Kaiki Taro Inoue}
\address{Faculty of Science and Engineering, 
Kindai University, Higashi-Osaka, 577-8502, Japan}
\begin{abstract}
By monitoring a large number of stars in the Local Group galaxies, 
we can detect nanolensing events by sub-lunar mass compact objects (SULCOs) such as primordial black holes
(PBHs) and rogue (free-floating) dwarf planets in the Milky Way halo.
In contarst to microlensing by stellar-mass objects, the finite-source size effect
 becomes important and the lensing time duration
 becomes shorter ($\sim 10^{1-4}\,\textrm{s}$).
Using stars with $V<26$ in M33 as sources, for one-night observation, 
we would be able to detect
$10^{3-4}$ nanolensing events caused by SULCOs in the Milky Way halo with a mass of 
$10^{-9}\ms$ to $10^{-7}\ms$ for sources with S/N$>5$ if SULCOs
constitute all the dark matter components. Moreover,
we expect $10^{1-2}$ events in which bright blue stars with S/N$>100$ are weakly
amplified due to lensing by SULCOs with a mass range of 
$10^{-11}\ms$ to $10^{-9}\ms$. Thus the method 
would open a new window on SULCOs in the Milky 
Way halo that would otherwise not be observable.
\end{abstract}
\end{frontmatter}


\section{Introduction}
Currently, there has not been stringent observational 
constraint on the abundance of SUb-Lunar mass Compact 
Objects (SULCOs) with a mass of $10^{-13}\ms \le M \le10^{-9}\ms$ as 
the dark matter candidates\citep{carr16}. SULCOs can be either small planets,
satellites, or primordial black holes (PBHs)\citep{carr74, carr75,
hawking82, yokoyama97, inoue03}. 
Microlensing tests such as the MACHO and EROS 
collaborations ruled out the possibility that the compact objects with
a mass of $10^{-7}M_{\odot}\le  M\le  10^{-3}M_{\odot} $
constitute the Milky Way halo\citep{paczynski86, alcock97, alcock00, tisserand07}.
On the other hand, femtolensing test of Gamma-ray bursts (GRBs) ruled
out the mass range $10^{-16}M_{\odot}\le M\le
10^{-13}M_{\odot}$ assuming that the GRBs are at cosmological
distance so that the angular size of the GRB source is 
sufficiently small~\citep{marani99, barnacka12}. The SULCOs also induce 
picolensing of GRBs but the observational limit 
is very weak~\citep{marani99}. The dynamical constraint
on the amount of SULCOs in the dark halo is also less stringent\citep{carr99,carr16}.
Recent lensing analyses based on the time variability of stars in the Milky Way disk
using the data from the Kepler satellite, have yielded a constraint on the mass range 
$2 \times 10^{-9}\ms \le  M \le 10^{-7}\ms$\citep{griest11,griest13,griest14}. However, the constraint
is limited to local SULCOs at a distance $<4 \,\tr{kpc}$. Since the size of
the Milky Way halo is much larger, it is important to constrain
the abundance of SULCOs that reside at distance $>4\, \tr{kpc}$ as well.   
The neutron-star capture constraint for a mass range of $10^{-15}\ms \le 
M \le 10^{-9}\ms$ \citep{capera13} depends 
on the assumption that the PBHs reside in globular clusters, thus the
limit is uncertain\citep{carr16}.

SULCOs may be free-floating or rogue dwarf planets that have been ejected
from developing or developed planetary systems\citep{ma16, smullen16}. 
Although the mass scale of the detected rogue planets\citep{penaramirez16}, typically a Jupiter mass scale, 
are much larger than the SULCO mass scale, a large
number of rogue dwarf planets may reside in the Milky
Way halo. For instance, the gravitational 
perturbation by passing stars may cause destruction of 
extrasolar planetary systems that may correspond to 
the Kuiper belt objects or the Oort cloud comets.     

In this paper, we propose a method using gravitational
lensing to constrain the abundance of SULCOs in the Milky Way halo: By monitoring a large number 
of individual bright stars in the Local Group galaxies such as M33\citep{abe00}, we can obtain
stringent constraint on the abundance of SULCOs as they induce
amplification of the background source stars. 
The key factors are
the source size and the scale of lensing time duration. In order to
constrain compact objects with a sub-lunar mass via gravitational lensing, 
we need to have sources
that are more distant than LMC or SMC for which the angular source size 
is smaller than the angular Einstein radius. Since the Einstein angular radius
of SULCOs (typically $\lesssim 10^{-9}$ arcsec) is comparable to
the radius of source stars, we need to consider the
finite-source size effect \citep{witt95}.
Even if the angular source size is larger than the Einstein angular radius of 
SULCOs, we can still detect the weak amplification of source stars if they are
bright enough. Note that the finite-source size effect and the feasibility of
observation with short time duration have not been explored in \citet{abe00}.
Moreover, the scale of lensing time duration becomes shorter as the lens mass decreases. Thus it is
important to assess whether currently available telescopes can probe
SULCOs with a reasonable observation time. In what follows, for simplicity, we
assume that the surface brightness of source stars is constant (limb
darkening is not taken into account) and circular. 
\section{Lensing by SULCOs}
\subsection{Significance and amplification}
If a source star is bright enough, the detectable
lensing amplification can be weak: the relation between the
signal-to-noise ratio for an unlensed source $\eta$ and that for an amplified event $\xi$ (an excess in flux) can be estimated as follows.

First, the signal-to-noise ratio $\eta$ is written in terms of
photon counts for the signal $S$ and those for the 
sky background and the other noise sources $N$ 
during an exposure time $T_e$,
\BE
\eta=\f{S}{\sqrt{S+N}}. \label{eq:1}
\EE
If the source is amplified by an amplification factor $A$, 
then the photon counts for the signal during an exposure time 
$T_e$ is increased to $S+\Delta S= A S$. The 
signal-to-noise ratio $\xi$ for an amplified event is then given by 
\BE
\xi=\f{\Delta S}{\sqrt{S+\Delta S+N}}. \label{eq:2}
\EE
Using equations (\ref{eq:1}) and (\ref{eq:2}), 
the amplification factor $A$ is given by
\BE
A(\eta,\xi,S)=1+\f{\xi^2}{2 S} + \xi \sqrt{
              \f{1}{\eta^2} + \f{\xi^2}{4 S^2}}.
               \label{eq:3}
\EE
In what follows, we assume that the photon counts of unlensed and lensed 
source are sufficiently large, i.e., $\eta^2 \ll S$ and $\xi^2 \ll S$. 
Then, equation (\ref{eq:3}) is reduced to a simpler form $A=1+\xi/\eta$.
For instance, if we would like to observe an amplified event at
$5\,\sigma$ and the significance of the unlensed source is 
$5\, \sigma$, then the lensing amplification should satisfy $A>2$ and 
the event corresponds to a strong lensing. If the signal-to-noise 
ratio for an unlensed star is $\eta=20$, 
then for detecting an event with $\xi=5$, the lensing amplification 
should satisfy $A>1.25$ and the event corresponds to a weak or strong 
lensing. If the significance of an unlensed star is $\ll 5 \sigma$, then the
amplification should satisfy $A \gg 2$, which corresponds to a 'very
strong' lensing.
\subsection{Maximum distance to lens}
Since the distance to the source stars at galaxies in the Local Group is 
much smaller than the present Hubble length, we can neglect the
effects of cosmological expansion. Hence, we omit a term ''angular diameter''
in the following. 
   
If the distance to the lens is too large, the angular Einstein radius
$\theta_E$ is so small that the amplification of an extended source 
cannot be observed. In what follows, we derive the maximum 
distance $D_{\max}$ to the lens
for which the amplification by SULCOs is detectable with a given 
signal-to-noise ratio $\xi_0$.

First, we consider the case in which the angular Einstein radius
$\theta_E$ is equal to or smaller than the angular size of 
the unlensed source star $\theta_a$, i.e., $\epsilon \equiv
\theta_E/\theta_a \le 1$. In this case, we expect a weak amplification
of a source star due to the finite-source size effect. The amplification factor $A$ is then 
approximately given by
\BE
A \approx \sqrt{1+4\epsilon^2}, \label{eq:4}
\EE
provided that a disk with radius
$\theta_E$\citep{witt94} centered at a point mass 
is totally contained in the source star (the angular source-lens
separation is sufficiently smaller than the angular source radius).  As shown in figure 1, 
equation (\ref{eq:4}) approximately holds even if $\epsilon \sim 1$.
The error is typically less than a few percent.
\begin{figure}[h]
\begin{center}
\includegraphics[width=102mm]{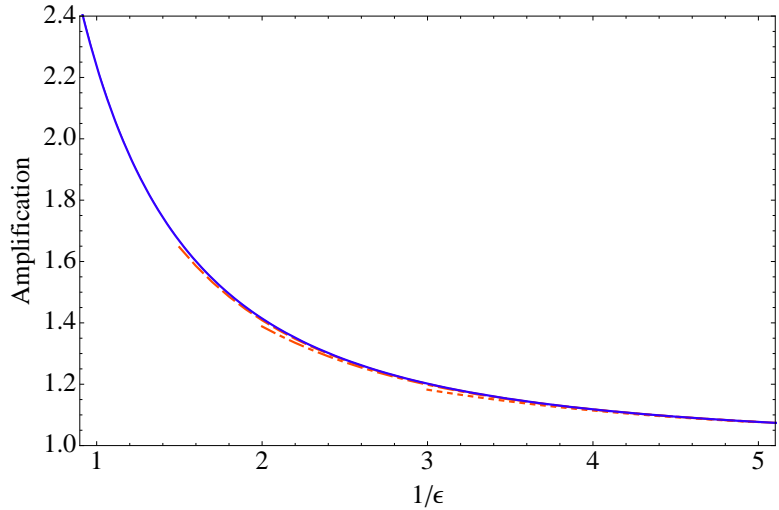}
\caption{Amplification by a point mass as a function of source
 radius. The surface brightness profile of a source is assumed to be a top-hat type. 
$1/\epsilon=\theta_a/\theta_E$ is the source radius in unit of
 an Einstein radius. The approximated amplification factor
in equation (4) is plotted as a blue solid curve. The exact amplifications for
a circular top-hat sources centered at $(\theta_E/2,0),
 (\theta_E,0), (2\theta_E,0)$ at the source plane are
plotted as dashed, dot-dashed and dotted orange curves,
 respectively. A point mass is placed at the center of coordinates.}
\label{f0}
\end{center}
\end{figure}
From equations (\ref{eq:3}) and (\ref{eq:4}),
we can calculate the minimum value for 
the size ratio $\epsilon_{\min}$ as a function of 
a given signal-to-noise ratio $\xi_0$ 
for amplification and that for the source star $\eta$.  
Assuming $\eta^2 \ll S$ and $\xi^2 \ll S$, for a given $\eta$, 
the minimum of $\epsilon$ is given by
\BE
\epsilon_{\min}= \f{\xi_0}{2 \eta} 
\sqrt{1+\f{2 \eta}{\xi_0}}. \label{eq:5}
\EE
The angular Einstein radius $\theta_E$ for a point mass
$M$ can be written in terms of the distances to the lens $D_L$, 
to the source $D_S$, and between
the source and the lens $D_{LS}$ as
\BE
\theta_E=\sqrt{\f{4 G M}{c^2}\f{D_{LS}}{D_L D_S}},
\label{eq:thetaE}
\EE 
where $c$ is the light velocity and $G$ is the gravitational 
constant. Then the size ratio is given by
\BEA
\epsilon&=&0.56\times \biggl(\f{D_S}{840 \, \textrm{kpc}}  \biggr)^{1/2}
\biggl(\f{a}{R_{\odot}}  \biggr)^{-1}
\biggl(\f{M}{10^{-9} M_{\odot}}  \biggr)^{1/2}
\nonumber
\\
&\times& \biggl(\f{1-x}{x}  \biggr)^{1/2} \label{eq:epsilon}
\EEA
where $a$ is the radius of the source star 
and $x\equiv D_L/D_S$ is the normalized distance to the lens.
For a given radius $a$ and a mass $M$, the size ratio $\epsilon$
is a monotonically increasing function of $D_S$ and  
a decreasing function of $x$. For neighbouring lenses at $x \ll 1$, we have 
$x \propto (\epsilon a)^{-2}$. Using equation (\ref{eq:5}) and equation (\ref{eq:epsilon}),
we can calculate the maximum distance $D_{L ,\max}$ ($x_{\max}$ if
normalized by $D_S$) 
for which one can detect a lensing event for a given $\xi_0$ and $\eta$.

Second, we consider the case in which $\theta_E$ is larger than $\theta_a$, 
i.e., $\epsilon > 1$. In this case, we expect a strong
amplification of a source star if the source is totally contained in a
disk with $\theta_E$ (small source-lens separation). The mean amplification is 
$\langle A \rangle  \gtrsim \sqrt{5} \approx 2.2$ if averaged over the
source position inside $\theta_E$.  For a given $\xi_0$, we can
calculate $\eta$ that gives $A \gtrsim 2.2$. Then, equation
(\ref{eq:epsilon}) gives the range of $x$ for which $\epsilon>1$ is
satisfied. If the source star is not resolvable, i.e., $\eta \ll 5$,
then the magnification factor should be large enough: $A\gg 2 $. In this
case (pixel lensing), the source star should be sufficiently close to
the lens that $A>1+5/\eta$. 
 
\subsection{Optical depth}
First, we consider lensing events with weak amplification ($\epsilon \le
1$) due to the finite-source size effect.
For these events, the amplification factor 
is almost constant when a disk with a radius of $\theta_E$ centered at a point mass 
is totally contained in the source (small source-lens separation) but it suddenly drops to unity if the
disk passes through the source star. Therefore, the lensing cross
section (defined as the total area of a source that yields a constant 
amplification) is $\sim \pi (D_L \theta_a)^2 $. Then the lensing 
optical depth (defined as the total area of sources that yields a constant 
amplification divided by the whole target area) for an extended source
with $\epsilon \le 1 $ is approximately given by
\BE
\tau(\epsilon \le 1)\approx \int_{D_{L}(\epsilon=1)}^{D_{L,\max}}
\f{\rho(D_L)}{M}\pi [\epsilon^{-1}\theta_E D_L]^2 dD_L, \label{eq:taus} 
\EE
where $D_L$ is the distance to the lens and
$\rho(D_L)$ denotes the density of dark matter at $D_L$.
Although equation (\ref{eq:taus}) gives
a slightly larger value for $\epsilon \sim 1$, for 
the purpose of our analysis, such a small contribution 
(typically less than a few percent) is negligible. From 
equation (\ref{eq:thetaE}) and equation (\ref{eq:taus}),
we have 
\BE
\tau(\epsilon \le 1)\approx \f{4 \pi G}{c^2} D_S^2 
  \int_{x(\epsilon=1)}^{x_{\max}} 
\f{\rho(x) x(1-x)} {\epsilon^2(x)}    dx, 
\label{eq:taus2} 
\EE
where $x_{\max}=D_{L,\max}/D_S$. 

Second, we consider lensing events with strong amplification ($\epsilon
> 1$), in which the source size is sufficiently 
smaller than the Einstein radius and the source star is resolvable
($\eta \ge 5$). Then the 
lensing cross section for a point mass\footnote{To be more precise, the cross section should be 
$ \sim 5\pi (D_L \theta_E)^2/(1+\xi/\eta)^2 $. For $\eta=\xi=5$, 
it becomes $ \sim \pi (D_L \theta_E)^2$.} is $ \sim \pi (D_L
\theta_E)^2$ for 
which the mean amplification is $\langle A \rangle =\sqrt{5} \approx 2.2$.
The lensing optical depth is given by 
\BE
\tau(\epsilon>1)\approx \f{4 \pi G }{c^2} D_S^2 
  \int_{0}^{x(\epsilon=1)} 
\rho(x) x(1-x)    dx.
\EE
If the source star is unresolvable ($\eta \ll 5$), the cross
section for a point mass should be much smaller than $\pi
(D_L\theta_E)^2$, which depends on the detectable amplification factor $A=1+\xi/\eta$. 
\subsection{Duration time}
In terms of the transverse velocities $\vel_O$,  $\vel_L$, 
and $\vel_S$, of the observer, the lens at redshift $z_L$, 
and the source at redshift $z_S$ in the source plane,  
the effective transverse velocity in the 
source plane is given by
\BE
\vel_{\textrm{eff}}=\f{1}{1+z_S}\vel_S-\f{1}{1+z_L}\f{D_S}{D_L}
\vel_L+\f{1}{1+z_L}\f{D_{LS}}{D_L}\vel_O. \label{eq:10}
\EE
For lenses $D_L\ll D_S$ and $z_O\sim z_L\sim 0$, the root-mean-square
of the effective velocity averaged over the lens objects at $x\ll 1$ is
\BE
\sigma_{\textrm{eff}}(x)=\langle |\vel_{\textrm{eff}}(x)|^2 \rangle^{1/2} 
\approx x^{-1} \langle |\vel_L|^2 \rangle^{1/2}, \label{eq:11}
\EE
where we assumed that $v_S\sim v_L \sim v_O$.
The average lensing duration time $\Delta T(x)$ is equal to the
average width of a disk ($=\pi/2\times$radius) that gives a lensing cross
section divided by the standard deviation of the effective
transverse velocity $\sigma_{\textrm{eff}}$,
\BE
\Delta T(x) \approx \begin{cases}
\cfrac{\pi a}{2 \sigma_{\textrm{eff}}
(x)},~~\epsilon \le 1 \\
\cfrac{ \pi a_E(x) }{2 \sigma_{\textrm{eff}}
(x)},~~\epsilon>1,
\end{cases}
\label{eq:deltaT}
\EE
where $a_E(x)=D_S \theta_E(x)$ 
is the Einstein radius\footnote{To be more precise, $\theta_E(x)$ should
be replaced by $\sqrt{5}\theta_E(x)/(1+\xi/\eta)$. } of a lens at a distance $x$ measured projected 
onto the source plane.
\subsection{Event number}
Let us consider an observation that consists of 
$N_e$ consecutive shots with an exposure time $\Delta T_e$. Then 
the differential effective optical depth for a spatial interval from
$x$ to $x+dx$ is
\BE
d \tau_{\textrm{eff}} \approx 
N_e \Delta T_e \Delta T^{-1}(x) \tau' dx, 
\label{eq:dtaueff}
\EE 
where $\tau'\equiv d \tau /dx$. The condition for 
the detection of a lensing event $\Delta T_e \le \Delta T $ sets the lower limit for the 
distance to the lens $x_\textrm{min}$. 
The effective lensing optical depth for such an observation is given by 
\BEA
\tau_{\textrm{eff}}&\approx&
\f{8 G  D_S^2 N_e \Delta T_e}
{c^2 } \biggl[
\int_{x(\epsilon=1)}^{x_{\max}}
\f{\rho(x)x(1-x)\sigma_{\textrm{eff}}(x)}
{ a \epsilon^2(x)} dx
\nonumber
\\
&+&
2 
\int_{x_{\min}}^
{x(\epsilon=1)}
\f{\rho(x)x(1-x)\sigma_{\textrm{eff}}(x)}
{a_E(x)} dx\biggr].
\label{eq:tau}
\EEA
Plugging equation (\ref{eq:thetaE}) into equation (\ref{eq:tau}),
we have
\BEA
\tau_{\textrm{eff}}&\approx&
N_e \Delta T_e \biggl[
\f{2 a D_S }{M}
\int_{x(\epsilon=1)}^
{x_{\max}}
\rho(x) x^2 \sigma_{\textrm{eff}}(x) dx
\nonumber
\\
&+&
4 \sqrt{\f{G D_S^3}{M c^2}}
\int_{x_{\min}}^
{x_(\epsilon=1)}
\rho(x)\sqrt{x^3(1-x)}\sigma_{\textrm{eff}}(x) dx\biggr]. 
\nonumber
\\
\label{eq:taueff}
\EEA
Multiplying equation 
(\ref{eq:taueff}) by the total number of observable 
stars $N_*$ for an exposure time $\Delta T_e$ and a 
signal-to-noise ratio $\eta$ for source stars,  
we obtain the total expected event number
\BE
E(N_e,\Delta T_e,\eta,\xi_0)=N_*(\Delta T_e,\eta) 
\tau_{\textrm{eff}}(N_e,\Delta T_e,\eta,\xi_0).
\EE
\subsection{Order estimate}
To probe SULCOs by monitoring a number of background stars, 
we take into account the finite-source size effect: 
the Einstein radius of a point mass should be 
sufficiently large with respect to the radius of the source star. 
In other words, the lens should be sufficiently close to us.
However, in that case, the lensing duration time becomes small and thus 
the exposure time for each snapshot must be small. This 
means that the target source stars should be sufficiently bright.

Let us estimate the event rate of nanolensing more quantitatively. First,
we need a number of bright stars at a distant place, i.e., large $D_S$
in order to reduce the angular source size. For simplicity, we assume
that the source radius $a$ is constant (typically of the order of the solar
radius), and the smallest size ratio is $\epsilon=1$. This corresponds to
strong lensing events with an amplification factor $A\gtrsim 2$ and
$\epsilon \ge 1$. We also fix $D_S$
and assume that the lenses are sufficiently close to us, i.e., the
density of dark matter $\rho(x)$ is
constant in $x$, and the number of observable source is fixed. 
From equations (\ref{eq:taus2}), (\ref{eq:deltaT}) and (\ref{eq:dtaueff}), we have 
the distance to the lens $D_L \propto M$, 
the lensing duration time $\Delta T \propto D_L\propto M$ and the
optical depth $\tau \propto D^2_L\propto M^2$. Therefore, the event
number is proportional to $\tau/\Delta T\propto M$. Thus, the observable
distance to a strong lensing event, the duration time, and the event number are all
proportional to $M$. In real setting, the number of available 
source stars depends on the exposure time $\Delta T_e< \Delta T$. Therefore,
the event number depends on $M$ more strongly.

As a source galaxy, we adopt a spiral galaxy M33 at $\sim 840\,
\tr{kpc}$\citep{freedman91}.
Although M33 is more compact and less massive than M31, the flux contribution from faint
unresolved stars per pixel is expected to be smaller than M31 as M33 is 
a typical face-on galaxy with much diffuse spiral arm where bright A-
and F-type main sequence stars are relatively abundant. Therefore, for
the purpose of observing weak amplification of bright resolved stars, M33 would be
much suitable. The luminosity function of the whole region of the  
disk and halo is known only at the upper end
($V<19.2$)\citep{freedman85}. 
To estimate the luminosity function of stars with $V>19.2$ in M33, we assume that it has the same
profile of the local luminosity function of the Milky Way \citep{bahcall80} except for the
normalization. Note that the slope $0.677$ at $V=21$ is consistent with 
the observed value $0.67 \pm 0.03$\citep{freedman85}. To normalize the luminosity function, we use 2112 blue stars
($U-V<0$ and $U-B<0$) and 389 red giants ($B-V>1.8$), which are complete to $V=19.5$\citep{ivanov93}.
It turned out that the number of stars is $6\times 10^5$ for $19.5<V<23$ and
$3\times 10^7$ for $19.5<V<26$.   

The typical time duration scale $\Delta T$ is given by the velocity of a lens at
which the angular Einstein radius is equal to the source radius,
i.e., $\epsilon=1$. For instance, for M33, equation
(\ref{eq:epsilon}) gives that the typical distance to the 
lens with a mass of  $M=10^{-9}\,\ms$ is $D_L=60\,\tr{kpc}$ assuming that the
source radius is $a=2R_\odot$, i.e., the typical size of a A-type 
main-sequence star, which is common at the M33 disk. Then, assuming 
the standard deviation of velocity $\sigma_{\tr{eff}}=200\,\tr{km/s}$ for lenses that reside in
the halo of the Milky Way, equation (\ref{eq:deltaT}) gives 
$\Delta T \sim 10^3\, \tr{s}$ for $M=10^{-9}\,\ms$. Assuming
the local dark matter density $\rho_0=0.0079 \ms/\tr{pc}^3$, and the
dark matter density for the halo of the Milky Way \citep{alcock97,alcock00},
$\rho(r)=\rho_0(r+(5\,\tr{kpc})^2)
/(r^2+(5\,\tr{kpc})^2)$\footnote{We omitted the contribution from SULCOs
in the halo of M33 as the Einstein radii are much smaller than those in
the halo of the Milky Way.}, and the 
optical depth for an observation with exposure time $\Delta T_e=\Delta
T/2=5\times 10^2\,\tr{s}$ is $\tau=3\times 10^{-7}$.  
For a 8\,m class telescope, the V-band magnitude limit (S/N=5)
for an exposure time of $5\times 10^2\,\tr{s}$ is typically 
$V\sim 26$. Therefore, the expected lensing event number is $\sim 10$
provided that point masses with $M=10^{-9}\,\ms$ constitute all the dark
matter. For a total integration time $\sim 7\, \tr{hours}$ with a field-of-view of 
$1\,\tr{deg}^2$, we expect $100-1000$ events. If the lenses have a mass 
$M=10^{-11}\,\ms$, the typical time duration is $\sim 8\, \tr{s}$ and
the expected event number is $<10$ for a total integration time $\sim 7\, \tr{hours}$.  
These numbers give approximate lower limits of event number as we did
not take into account contributions from weakly amplified events ($\epsilon <1$) in this
order estimate. It should be noted that the contribution from neighbouring
unresolvable stars in each pixel is negligible because the surface
brightness of disk stars in M33 is typically $V \sim 26/\tr{arcsec}^2$
\citep{ibata07} at angular distances $\lesssim 1^\circ$, which is much
fainter than that of sky.
 
\subsection{Simulation of observation}
In order to assess the feasibility of our method, we assume an
observation of M33 with a 
wide field (1.5 degree in diameter) camera that is attached to
a 8\,m class telescope. For a limiting magnitude $V=26$, with a seeing of $\sim 0.6$
arcsec, it can resolve $\sim 10^7$
stars at angular distances $\theta>14$ arcmin from the center assuming 
that the surface brighteness is proportional to $\exp(-\theta)$ \citep{abe00}. 
We also assume a relatively good condition, such that 
seeing is $\sim 0.6$ arcsec and the sky is dark. 
We use the same Milky Way halo model used in subsection 2.6, and we also
take into account the size of stars, which depends on the
luminosity. We use the mass-radius relation 
for luminous main sequence A and B type stars ($2\ms \lesssim  M_* \lesssim 20\ms$), which yields 
$a/a_\odot=-(V-20)+8$ \citep{binney00}. As the
number of bright red stars are smaller (about 1/5) than that of bright
blue stars for $V<19.5$, we expect that the effect of red giants is negligible at the bright end. 
As shown in the right panel of figure 2, the event number for 
$(T_e,\eta)=(60\,\tr{s}, 5)$(red,dot-dashed) is $\sim 1000$ for 
$M=10^{-9}\,\ms$ to  $M=10^{-7}\,\ms$.
If the exposure time is $1\,\tr{s}$, the event number can be further
increased to $\sim 10000$ but if one considers
the CCD readout time of $\sim 30\,\tr{s}$, the event number is reduced
by more than 10 times in real setting (the left panel of figure 2)
For masses with $M< 10^{-10}\,\ms$, 
the lensing effect becomes weak due to the finite
source-size. However, if we restrict the source stars to only bright
ones, this difficulty can be somehow avoided (though the CCD saturation inhibits
the measurement of very bright stars $V\lesssim 20.5$). 
For instance, for the parameters
$(T_e,\eta)=(60\,\tr{s}, 100)$(red, dashed), we expect $10-100$ events
during an observing time of $7\,\tr{hours}$. This is due to the weak
lensing effect: Even if the source size is much larger than the Einstein
radius, the source can be slightly amplified by a lens. As one can see
in figure 3, the maximum lensing time duration
$\Delta T(x_{\tr{max}})$ and the maximum distance to the lens
$x_{\tr{max}}$ increase as the
signal-to-noise ratio $\eta$ increases as a result of small
$\epsilon_{\tr{min}}$. This boosts the chance of detection of amplified
events caused by SULCOs. Of course, it is challenging to detect such a small change as 
other systematics such as intrinsic variability (flare of stars) 
or the CCD noises can easily hamper such
detection. However, in principle, one can discriminate lensing
events by looking into the ``colourless'' feature in the time variability 
and by comparing the measured flux to the templates of predicted light curves.  
\begin{figure}[tbp]
\begin{center}
  \begin{tabular}{c}
\hspace{-0.8cm}
\begin{minipage}{0.5\hsize}
      \begin{center}
        \includegraphics[width=56mm]{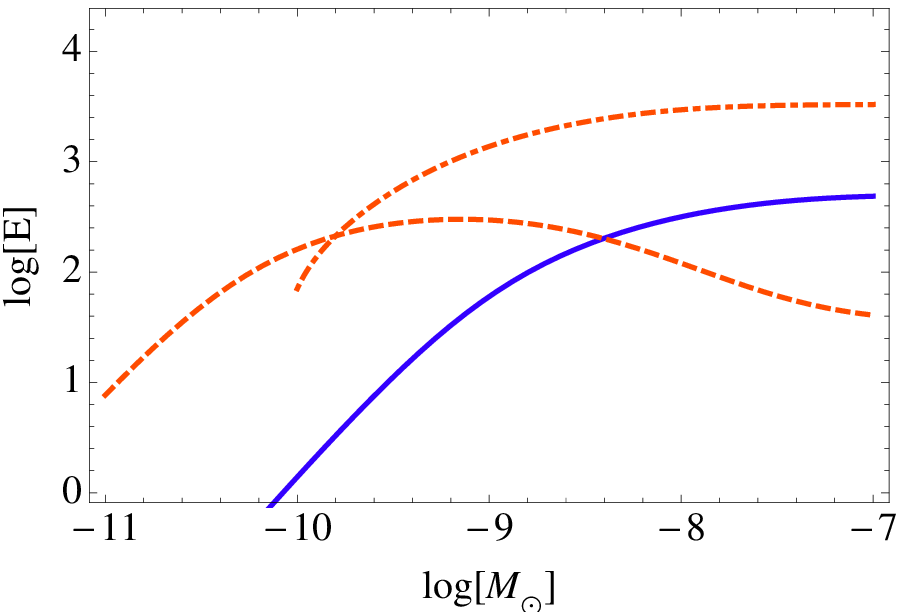}
      \end{center}
    \end{minipage}
    \begin{minipage}{0.5\hsize}
      \begin{center}
        \includegraphics[width=58mm]{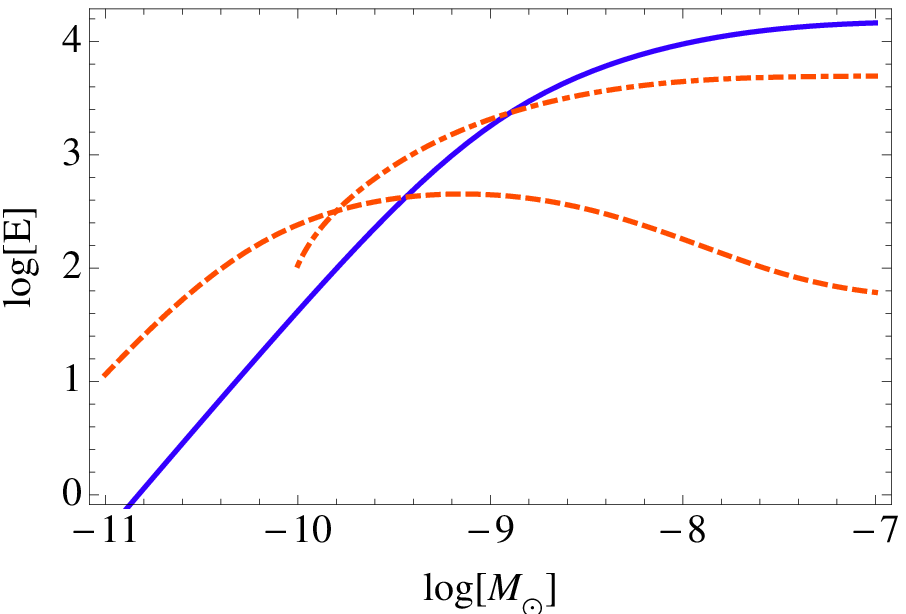}
      \end{center}
    \end{minipage}
   \end{tabular}
  \caption{Expected event number as a function of SULCO mass for $7\,\tr{hours}$
 observation of M33 using a wide field camera attached to a 
8\,m class telescope with (left) or 
without (right) the CCD readout time of $\sim 30\,\tr{s}$ taken into account.
The exposure time for each snapshot and the signal-to-noise
 ratio of the source stars are $(T_e,\eta)=(1\,\tr{s}, 5)$(blue, full curve), $(60\,\tr{s}, 5)$(red,
 dot-dashed), and $(60\,\tr{s}, 100)$(red, dashed). For the low-end
 ($M<10^{-10}\,\ms$),  the finite source-size effect becomes important.}
  \end{center}
\end{figure}
\begin{figure}[tbp]
\begin{center}
  \begin{tabular}{c}
\hspace{-0.8cm}
\begin{minipage}{0.5\hsize}
      \begin{center}
        \includegraphics[width=56mm]{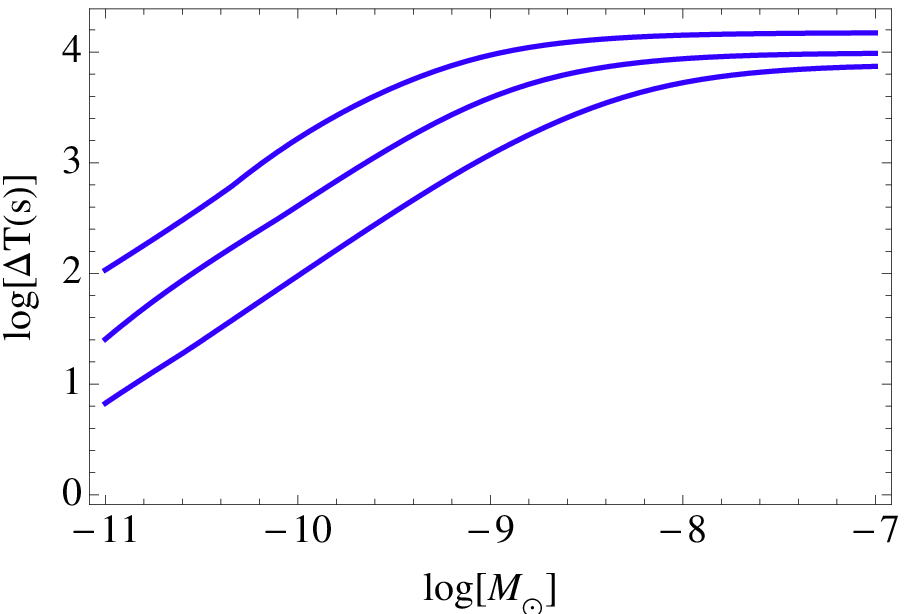}
      \end{center}
    \end{minipage}
    \begin{minipage}{0.5\hsize}
      \begin{center}
        \includegraphics[width=59mm]{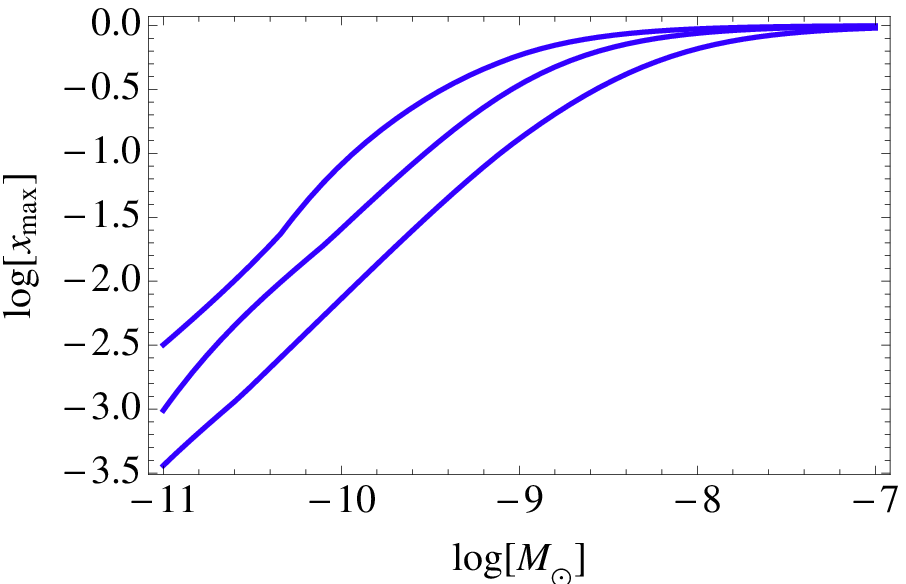}
      \end{center}
    \end{minipage}

  \end{tabular}
  \caption{The maximum lensing time duration (left) and the maximum
 distance to the lens (right) as a function of SULCO
 mass. The signal-to-noise ratios are 5,20,100 from the bottom to the top.}
  \end{center}
\end{figure}
\section{Conclusion and Discussion}
In this paper, we have discussed that nanolensing events 
by dark sub-lunar mass compact objects (SULCOs) such as PBHs 
and free-floating dwarf planets can be detected by
monitoring $10^{5-7}$ stars in the Local Group galaxies
such as M33. The typical lensing time scale is $\sim 100\,\tr{s}$.
The exposure time for each snapshot must be $\sim 100\,\tr{s}$, which
is the typical time scale of nanolensing variability for 
SULCOs with a mass of $\sim 10^{-9}\,\tr{M}$. Using a 8\,m class
telescope, $10^{3-4}$ events per night would be detected if
SULCOs constitute all the dark matter. Moreover,
we expect $10^{1-2}$ events from weak 
amplification of very bright stars caused by SULCOs with a mass range of 
$10^{-11}\ms$ to $10^{-7}\ms$ though detection of such change may be a
challenging task. Our method would provide a 
stringent constraint on the abundance of SULCOs at the distance 
$0.1-100\,\tr{kpc}$ from us.

As a source galaxy, we have considered M33 as it has a relatively 
large number of blue main sequence stars and much diffuse spiral arms
in comparison with M31. These features are important for observing 
weak amplification (which was not studied in \citet{abe00}) by SULCOs from time variability of very bright 
source stars. However, for the purpose
of detecting nanolensing events, M31 would be much suitable as the
number of available source stars is larger than M33 though the effect of
blending due to neighbouring stars may be stronger as it is not face-on.
For detecting SULCOs via weak amplification of source stars, other
galaxies at a farther distance would be suitable if 30-m class telescope
is available in the coming decade. 
 
We have taken into account the finite source-size effect, which is
important for estimating the weak amplification caused 
by SULCOs. However, the effect of spatial variability in the source brightness
such as limb darkening has not been taken into account.  Such an effect becomes important when the
impact parameter of the source is comparable to the radius of the star.
More precise treatment is necessary for the cases in which the angular
Einstein radius is approximately equal to the angular source size.  

A part of SULCOs may consist of free-floating dwarf planets.
Detection of these objects in the intergalactic space is a challenging
task. Like MACHOs, these small unbounded objects may constitute a large
portion of baryonic masses in the disk or halo of our galaxy. Our method 
would open a new window on these small objects in the Milky 
Way halo that would otherwise not be observable.


\bibliographystyle{mnras}
\bibliography{SULCO}

\end{document}